\documentclass[12pt]{article}
\usepackage{epsf, cite, amssymb}
\usepackage{epsfig}

\makeatletter
\renewcommand\section{\@startsection {section}{1}{\z@}%
                                   {-3.5ex \@plus -1ex \@minus -.2ex}
                                   {2.3ex \@plus.2ex}%
                                   {\normalfont\large\bfseries}}
\renewcommand\subsection{\@startsection{subsection}{2}{\z@}%
                                     {-3.25ex\@plus -1ex \@minus -.2ex}%
                                     {1.5ex \@plus .2ex}%
                                     {\normalfont\bfseries}}
\makeatother

\newcommand{\bea}{\begin{eqnarray}}
\newcommand{\eea}{\end{eqnarray}}
\newcommand{\be}{\begin{equation}}
\newcommand{\ee}{\end{equation}}

\newcommand{\g}{\gamma}

\newcommand{\e}{\epsilon}

\newcommand{\dd}{\delta}

\newcommand{\acom}[2]{{ \left\{ #1, #2 \right\} }}

\newcommand{\p}{\partial}

\def\acom#1#2{{ \left\{ #1, #2 \right\} }}

\newcommand{\f}{\psi}
\newcommand{\df}{\dot{\f}}

\newcommand{\C}[1]{$(\ref{#1})$}

\begin{document}
\begin{titlepage}

\begin{center}

{April 6, 2004}
\hfill                  hep-th/0404056

\hfill EFI-04-12

\vskip 2 cm
{\Large \bf Structure in Supersymmetric Yang-Mills Theory}\\
\vskip 1.25 cm { Savdeep Sethi\footnote{email address:
 sethi@theory.uchicago.edu}}\\
{\vskip 0.5cm  Enrico Fermi Institute, University of Chicago,
Chicago, IL
60637, USA\\}

\end{center}

\vskip 2 cm

\begin{abstract}
\baselineskip=18pt

We show that requiring sixteen supersymmetries in quantum mechanical
gauge theory implies the existence of a web of constrained
interactions. Contrary to conventional wisdom, these constraints
extend to arbitrary orders in the
momentum expansion.  

\end{abstract}

\end{titlepage}

\pagestyle{plain}
\baselineskip=19pt

\section{Introduction}

Supersymmetry is a truly remarkable symmetry. What is perhaps more
surprising is that we still do not understand the full extent of the
constraints imposed by supersymmetry on  field theory and string
theory. The goal of this paper
is to further unravel these constraints. We will consider maximally
supersymmetric Yang-Mills theory with $16$ real supersymmetries.

The simplest case to analyze is quantum-mechanical supersymmetric
Yang-Mills theory. This is the theory that describes the low-energy
dynamics of D$0$-branes in type IIA string theory. In the past, the
techniques used to prove non-renormalization results in this theory
have generalized to both higher-dimensional field theory and string
theory. We can hope that the same will be true of this analysis.   

Our interest is in
the effective action describing the physics on the Coulomb branch of
this theory. Again for simplicity, let us restrict to the rank $1$
case; for example, $SU(2)$ broken to $U(1)$.  
The effective action can then be thought of as describing the dynamics of a
supersymmetric particle in $9$ dimensions. The position of the
particle is determined by $9$ scalar fields, $x^i(t)$, whose
superpartners are $16$ fermions, $\f_a(t)$.   

The effective action involves couplings constructed from $(x^i, \f_a)$
and derivatives of these fields. To any coupling, we can assign an order, denoted $n$,
given by 
\be
n = n_\partial + {1\over 2} n_f,
\ee
where $n_\p$ is the number of derivatives while $n_f$ is the number of
fermions. The order measures the relevance of the coupling at
low-energies. Terms with more derivatives are less relevant at low-energies. 

Because of the freedom to perform field redefinitions, the form of the
effective action is ambiguous. However, as we will show in
section~\ref{simplify}, there is a particularly nice choice of
fields. In terms of these fields, at the lowest order with 
$n=2$, the action is unique taking the free-particle form~\cite{Paban:1998ea}
\be \label{free}
S_1 = \int dt \, {1\over 2}\left( v^2 +  i \f \df \right), 
\ee
where $v= \p_t x$. All of the remaining couplings in the effective action are
constructed in terms of $(v,\f)$ only with no higher time derivatives. This
generalizes an observation of~\cite{okawaobs}\ employed
in~\cite{Kazama:2002jm}\ in a study of the $O(v^4)$ terms. This
simplification is special to quantum mechanics, although it should
have some analogue in higher-dimensional field theories.

\begin{figure}[ht]
\begin{center}
\includegraphics[width=16cm]{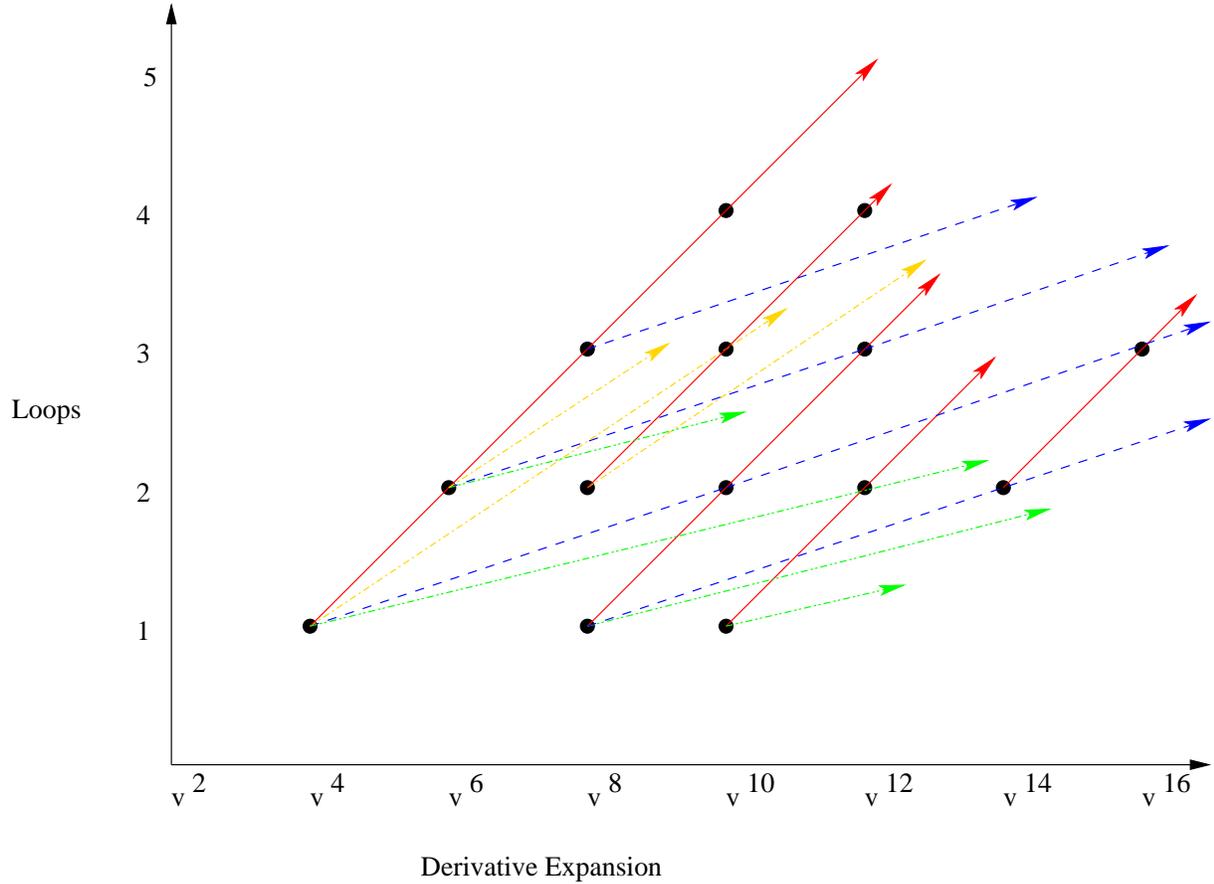}
\caption{A depiction of the web of induced couplings.}
\label{fig}
\end{center}
\end{figure}

At low orders in the derivative expansion, there are
non-renormalization theorems. Terms of $O(v^4)$ are only generated at
$1$-loop~\cite{Paban:1998ea, Hyun:1999hf, Kazama:2002jm}, while terms
of $O(v^6)$ are only generated at
$2$-loops~\cite{Paban:1998qy}. There are no known results beyond $O(v^6)$. The
usual intuition associated to this breakdown at $O(v^8)$
is the (heuristic) argument that these terms involve integrals over
all of superspace, while $O(v^4)$ and $O(v^6)$ terms only involve integrals
over a fraction of superspace. The latter interactions are therefore
special. This argument is heuristic because
there is no superspace construction that keeps manifest all $16$ real
supersymmetries. 
What we will find is that this intuition is actually 
incorrect. To some extent, we already knew this from an analysis of
higher rank theories~\cite{Sethi:1999qv,Dine:1999jq}, but we will find that this is true even
for rank $1$.

Our results are most easily explained via
figure~\ref{fig}.\footnote{This figure is best seen with a color viewer!} What we
will primarily study is a particular spin-spin coupling at $O(v^{2n})$, 
\be \label{spincoupling}
g_4^{(n,1)}(r) v^{2n-2} x_ix_j (\f \g^{ik} \f)(\f \g^{jk} \f). 
\ee   
In principle, this coupling can be generated at any order in
perturbation theory. Supersymmetry, however, imposes a much more rigid
structure: this coupling is never generated at $1$-loop, but can
be generated at higher loops. In terms of figure~\ref{fig}, this
spin-spin coupling is  only generated at points through which some ray
passes. It is
never generated at any anchor point. An anchor point is a point which
only originates rays. 
{}For example, consider the anchor located at coordinates
$(v^4,1)$. The existence of this point tells us that $O(v^4)$ terms
are generated at $1$-loop. Since this is an anchor point, the spin-spin
coupling given in~\C{spincoupling}\ is not generated. This is in accord with
both an explicit computation~\cite{McArthur:1998gs}\ at $1$-loop, and prior
supersymmetry arguments~\cite{Hyun:1999hf, Kazama:2002jm,
  Nicolai:2000ht}. 
However, the existence of $1$-loop
$O(v^4)$ terms generates a line of slope $1$ in the figure. The next
point on the line is at $(v^6,2)$. A $2$-loop spin-spin coupling is induced
at this order with a coefficient that is predicted in terms of the
$O(v^4)$ interactions. More precisely, in terms of the coefficient of
the $v^4 /r^7$ interaction.  Effectively, the $O(v^6)$ spin-spin coupling is
sourced by the $O(v^4)$
terms; had they been absent, there would be no $O(v^6)$ spin-spin
coupling. This is not particularly remarkable because we already know
that all terms at $O(v^6)$ are sourced by the $O(v^4)$ interactions~\cite{Paban:1998qy}. 

This gets much more interesting when we go to the next point on the
line at $(v^8,3)$. A $3$-loop exact spin-spin coupling is induced at this
order which corresponds to, 
\be
g_2^{(4,1)} \sim {1\over r^{25}}.
\ee
 This is regardless of whether any other coupling, for
example the $v^8$ interaction, is
generated at other orders in perturbation theory. Indeed, we can
predict the numerical coefficient of this spin-spin coupling from our knowledge
of the $v^4$ and $v^6$ interactions. Continuing along the line, we see
that an $O(v^{10})$ spin-spin coupling is induced at $4$-loops. The
coefficient for this interaction can be predicted in terms of the
coefficient of the $3$-loop contribution to the $v^8$ interaction, and
so on. 

This brings us to a critical open issue. Let us
return to the terms of $O(v^8)$. 
We will argue that a particular coupling is determined exactly at this
order; namely, the spin-spin coupling~\C{spincoupling}.  It is also
likely that there are additional couplings that can be determined
using similar arguments. Using the notation described in
Appendix~\ref{fermionstructures}, the $12$ fermion coupling
\be \label{can3}
 v_i v_j \ast \left\{ (\f \g^{ik} \f)(\f \g^{jk} \f) \right\},  
\ee
and the couplings~\C{c1}\ and \C{can2}\ are natural candidates. 

Are these couplings, in
conjunction with symmetry,
sufficient to completely determine all $3$-loop terms at $O(v^8)$?
If so, this would mean that there is no arbitrary $3$-loop solution to
the supersymmetry constraints 
at $O(v^8)$. We can hope that the answer is yes, but demonstrating
this will require understanding the full conditions imposed by
invariance under
supersymmetry; perhaps, including closure of the supersymmetry
algebra. Because our nice choice of fields enormously
simplifies the form of the effective action, it appears to me that
this question  can now be
fully answered for this theory.   

Returning to figure~\ref{fig}, we see that there are other lines of
slope $1$. Each of these lines is induced by the existence of the $1$-loop
$O(v^4)$ terms. 
If there is any other non-zero $v^{2n}$ interaction
then a ray of slope $1$ will emanate from that point on the
diagram. For example,  let us suppose there are non-vanishing $1$ and $2$-loop $v^8$
terms as depicted in figure~\ref{fig}. These terms are generically
expected to be non-vanishing, and in fact, the 2-loop term is known to
be  non-vanishing~\cite{Becker:1998cp}. They then correspond to new anchor points in the
diagram from which rays of slope $1$ extend. Since they are anchor points,
there is no $1$-loop or $2$-loop spin-spin coupling at $O(v^8)$, but there
are induced $2$-loop
and $3$-loop contributions to the spin-spin coupling at $O(v^{10})$ whose
coefficients can be predicted. 

Further, the new
anchor points give rise to new lines with slopes less than $1$. This
comes about in the following way. The $O(v^{2n})$ terms generate
corrections to the supersymmetry transformations (of the same loop
order) that allow you to
move $(n-1)$ steps in the derivative expansion. So the $O(v^4)$ corrections
connect terms in the derivative expansion that differ by $O(v^2)$, or
$1$ step. The
$O(v^8)$ terms, however, connect terms in the derivative expansion
that differ by $O(v^{6})$, or $3$ steps. Therefore, the $1$-loop
$O(v^8)$ term generates a $2$-loop $O(v^{14})$ term. Hence, there are lines of
slope $1/3$ in the diagram. On the other hand, the $2$-loop
$O(v^8)$ term generates a $4$-loop $O(v^{14})$ term which explains the
lines of slope $2/3$. Lastly, the $1$-loop $O(v^{10})$ anchor point
(should those terms be non-vanishing as we expect)
gives rise to lines of slope $1/4$ for exactly the same reasons. These
rays with different slopes extend from every node in figure~\ref{fig}\ although, for
clarity, only some of
the rays are actually depicted.

{}From the perspective of a perturbative field theorist, this web of
interactions must look bewilderingly complicated.  With the aid of
symmetry, however, we will see that the structure has remarkably simple origins.  
In fact, there is more structure than a single determined spin-spin
coupling. The most studied terms in the effective action have the
form, 
\be
g_0^{( n)} (r) v^{2n} + g_2^{( n)}(r) v^{2n-2} x^i v^j \f \g^{ij} \f + \ldots,
\ee
where $g_2^{( n)}(r)$ is the coefficient of the spin coupling. This
spin coupling at $O(v^4)$ has been studied in~\cite{Kraus:1998st,
Morales:1998dz, Serone:1998pg, McArthur:1998gs, Hyun:1999qd}. For
general $n$, we do not yet know how to separately fix $g_0^{( n)}$ and
$g_2^{( n)}$. However,
supersymmetry does fix the combination
\be
g_2^{( n)} + {i \{ g_0^{( n)} \}'\over 2 r}, 
\ee
in terms of more relevant interactions, just like the spin-spin
coupling. This is even true for anchor
points for which, 
\be g_2^{( n)} + {i \{ g_0^{( n)} \}'\over 2 r} =0, \ee
and the spin coupling is fixed in terms of the $v^{2n}$
interaction. There are plenty of relations of this sort between
interactions in the effective action.

There are many directions to explore. As I stressed earlier, the
complete set of supersymmetry constraints can, in
principle, now be determined for this theory. The extension to higher
rank, higher
dimensions, and less supersymmetry will involve novel issues, and
hopefully provide novel
results. For example, it is worth stressing that these results are
actually 
non-perturbative. When applied to Yang-Mills in three dimensions or
type IIB string theory, we
should be able to learn about instanton corrections to special
interactions~\cite{Paban:1998mp, Hyun:1998qf, Hyun:1999hm,
  Green:1998by, Sinha:2002zr}. The Matrix theory~\cite{Banks:1997vh}\ interpretation
of these relations should teach us something about M theory.   
Lastly, extending this analysis to 
N=4 Yang-Mills in four
dimensions should also prove interesting in light of the recent
conjecture about the structure of
certain perturbative gluon ampitudes~\cite{Witten:2003nn}.

\section{D0-Brane Dynamics}

\subsection{Some preliminaries}

The low-energy degrees of freedom describing the dynamics of D0-branes
consist of $9$
bosons, $x^i$, transforming in the vector representation of the
R-symmetry group $Spin(9)$. In addition, there are $16$ real fermions,
$\f_a$, transforming in the spinor representation. The bosons have
mass dimension $[x] =1$ while the fermions have mass dimension $[\f] =
3/2$. The Yang-Mills
coupling constant, $g^2$, has mass dimension $3$.

The effective action for these degrees of freedom takes the schematic form,
\be
S = \int dt \left( \, f_1(r) v^2 + f_2(r) v^4 + f_3(r)v^6 +
\ldots\right),
\ee
where $f_k v^{2k}$ simply represents all possible terms in the momentum
expansion of appropriate order.  We define the
$Spin(9)$ gamma matrices by the relation:
\be
\acom{\g^i}{\g^j} = 2\delta^{ij}.
\ee
The supersymmetry transformations can then be expressed in the
form,
\bea \label{susyalg}
\dd x^i &=& -i\e \g^i \f + \e N^i \f \cr
\dd \f_a & =& \left(\g^i v^i \e \right)_a + \left(M \e \right)_a.
\eea
We have lumped all the complicated corrections to the free-particle
supersymmetry transformations into $N$ and $M$.

What is known about the effective action can be summarized
as follows: at order
$2$, the action is unique taking the free-particle form~\C{free}. 
At order $4$, the action takes the form
\be
S_2 = \int dt \, \left( f_2^{(0)}(r) v^4 + \ldots f_2^{(8)}(r) \f^8
\right).
\ee
Again, the notation is schematic, and includes interactions involving
accelerations $a$, $\dot{a}$, $\df$, $ \ddot{\f}$ terms etc.
The technique developed in~\cite{Paban:1998ea}\ tells us that the $f_2^{(8)}(r)
\f^8$ interactions are $1$-loop exact. In a series of
papers~\cite{Hyun:1999hf, Hyun:1999qd, Kazama:2001sq, Kazama:2001gg, Kazama:2001bb},
culminating in~\cite{Kazama:2002jm}, this result was shown to imply the
non-renormalization of all other terms at this order, as conjectured
in~\cite{Paban:1998ea}.

Lastly, at order $6$, the argument of~\cite{Paban:1998qy}\ shows that the
$f_3^{(12)}(r) \f^{12}$ interactions are 2-loop exact. These
interactions are really determined by the terms in $S_2$. One way to
view this result is that the $\f^{12}$ interactions are slaves of the
$\f^{8}$ terms needed to obtain a closed supersymmetry algebra. Although
it has yet to be explicitly checked, we believe that all terms at
order $6$ are related by supersymmetry to the $f_3^{(12)}(r) \f^{12}$
interactions, and are therefore 2-loop exact. In fact, this will
become clear in light of our subsequent analysis. 

The argument given in~\cite{Paban:1998ea}\ breaks down at order $8$ for a simple
reason. Consider the `top form' interaction,
\be
f_4(r) \f^{16},
\ee
and vary $f_4$. At order $4$ and order $6$, the analogous variation of
$f_2^{(8)}(r) \f^8$ and $f_3^{(12)}(r) \f^{12}$ leads to a
non-vanishing $9$ fermion and $13$ fermion term, respectively. Since
these variations involve no space-time derivatives, they must either be
cancelled or vanish. This argument leads to the
non-renormalization results. However, in this
case, the variation automatically vanishes since we only have $16$
fermions. The order $8$ interactions are therefore not expected to be
special in anyway, and could in principle be generated at any order in
perturbation theory (and in higher dimensions, non-perturbatively).

\subsection{Simplifying the action}
\label{simplify}

Can we say more? To proceed, we first extend an observation
of~\cite{okawaobs, Kazama:2002jm}. 
Consider all terms of order $2n$ where $n>1$. Integrating by parts
allows us to express the action in a special form:
\be
S_n  = \int \left( f_n(x,v,\f) + a^i k_i + \df_a h^a \right).
\ee
Here $f_n$ contains no accelerations or higher time derivatives of
$x$, and no time derivatives acting on $\f$. All such terms are lumped
into $k_i$ and $h^a$.

Noting the special form of $S_1$ given in \C{free}, we see that the
field redefinition
\bea
x^i &\rightarrow & x^i + k^i,  \\
\f_a & \rightarrow & \f_a + {i\over 2} h_a,  \\
\eea
removes the acceleration and $\df$ terms while leaving terms of order
less than $2n$ invariant. By induction, we can remove all acceleration
and $\df$ terms leaving an action, aside from $S_1$, which
depends only on $x,v,\f$.

This is a nice simplification which teaches us that with this choice
of fields, the action has the
schematic form
\be \label{notmom}
S = \int dt \left( \, g_0(v,x) + g_2(v,x) x^i v^j \f \g^{ij} \f +
\ldots + g_{16}(v,x) \f^{16} \, \right),
\ee
where the scalar functions, $g_{m}$, depend on the $Spin(9)$ invariants
$(v^2, \, x^2, \,
x\cdot v)$. Note that the action \C{notmom}\ is not a momentum
expansion! Each $g_{2m}$ appears with some combination of $2m$ fermions, but
contains terms of all orders in velocity. There are unique
$g_2$ and $g_{16}$ functions because there are unique non-vanishing
$2$ fermion and $16$ fermion structures. For other choices of $m$,
there can be many
independent $2m$ fermion structures each appearing with its own $g_{2m}^{(i)}$
function. We will try to avoid delving into those details until later.

As a final simplification, we can choose each $g_{2m}$ to depend only on
$(v^2, \, x^2)$ and not on $(x\cdot v)$. 
The argument goes as follows: consider any term of the form
$$  g_{2m} (v^2, r) (x\cdot v)^k \left( x^{i_1} \ldots x^{i_l} v^{j_1}
\ldots v^{j_p} \right) T_{i_1 \ldots i_l , j_1\ldots j_p}^{a_1 \ldots
  a_{2m}} \f_{a_1} \ldots \f_{a_{2m}} $$
where $T$ is some structure constructed from $\g$ matrices.
Up to the introduction of acceleration and $v^2$ terms, we can make the substitution
$$ g_{2m} (v^2, r) (x\cdot v)^k \, \rightarrow \,  {1\over 2} \p_t
\{ \, \widetilde{g}_{2m} x^2 (x\cdot v)^{k-1} \, \}, $$
where we choose $ \widetilde{g}_{2m}$ to satisfy
\be \label{definetilde}
\left( 1+ {1\over 2 } r \p_r \right)\widetilde{g}_{2m} = g_{2m}.
\ee
This equation is always solvable. After this substitution, we can
integrate by parts to leave only terms depending on $ (x\cdot
v)^{k-1}$, $a$, or $\df$. The $a$ and $\df$ terms can be field
redefined away, and the procedure repeated until only $v^2$ type terms
remain. This is not an essential simplification, but it does make the
algebra cleaner.

\subsection{Demonstrating non-genericity}

Let us begin by supposing that the theory consists only of the free
particle terms in $S_1$ and the terms of order $2n$ in $S_n$ so
\be
S = S_1 + S_n.
\ee
We will
treat all the intervening terms with orders less than $2n$ as sources,
but first we need to understand the homogeneous solution to the
supersymmetry constraints.

With the simplifications described in section~\ref{simplify}, we can
express the terms in $S_n$ in the form,
\be \label{firsttwo}
g_0^{( n)} (r) v^{2n} + g_2^{( n)}(r) v^{2n-2} x^i v^j \f \g^{ij} \f + \ldots.
\ee
These terms will induce corrections to the supersymmetry
transformations. Let us expand the supersymmetry generators $\dd_a$ in
a series
$$ \dd_a = \dd_a^1 + \dd_a^2 + \ldots $$
where the $\dd_a^n$ term is induced by terms of order $2n$. The
invariance condition then reduces to the statement that
$$ \dd^n S_1 + \dd^1 S_n = 0. $$
The terms from $\dd^n S_1$ are particularly nice
taking the form,
\be
 - a^i \e N^i\f  - i \df M \e,
\ee where $N$ and $M$ are defined in~\C{susyalg}. It is key that
these terms involve either $a$ or $\df$. The strategy then is to
vary \C{firsttwo}\ and separate out the $a,\df$ terms.

{}For the $1$ fermion terms in the variation of \C{firsttwo}, it is
an easy exercise to separate out the $(a, \df)$ terms from the
rest. The resulting equations imply that
\bea
g_2^{( n)} &=& - {i \{ g_0^{( n)} \}'\over 2 r}, \label{1ferm} \\
\e N^i_0 \f &=& (2n-2)i g_0^{( n)} v^{2n-4} v^i v^k (\e \g^k \f) + i g_0^{(
  n)}v^{2n-2} (\e \g^i \f), \label{hom1} \\
M_0 \e & = & (2n-1) g_0^{( n)} v^{2n-2} v^k \g^k \e. \label{hom2}
\eea
The subscript on $N$ and $M$ denotes the number of fermions in the
terms under consideration. When we need to distinguish terms in $(N,M)$
generated at $O(v^{2n})$, we will again use the superscript notation $(N^{(n)},
M^{(n)})$. 
The noteworthy features of~\C{hom1}\ and~\C{hom2}\ are that both $N_0$
and $M_0$  depend on only $1$
gamma matrix, and that $M_0$ has the same form as the free-particle
result~\C{susyalg}. 

To see something interesting, we need to consider the $4$ fermion
terms in $S_n$. As described in Appendix~\ref{fermionstructures},
there are three such structures which take the form \bea
\label{fourfermi} S_n &= \ldots & + \, g_4^{(n,1)}(r) v^{2n-2}
x_ix_j (\f \g^{ik} \f)(\f \g^{jk} \f) + g_4^{(n,2)}(r) v^{2n-4} v_i
v_j (\f \g^{ik} \f)(\f \g^{jk} \f) \cr && + \, g_4^{(n,3)}(r)
v^{2n-4} x_ix_j v_k v_l (\f \g^{ik} \f)(\f \g^{jl} \f) + \ldots.
\eea This looks messy, but the observation we need to make only
concerns  the $g_4^{(n,1)}$ term. We want to study the supersymmetry
variation of \C{fourfermi}\ into $3$ fermions, but this has a
piece that looks like
$$ 4 g_4^{(n,1)}(r)  v^{2n-2} x^i x^j v_p (\e \g^{pik} \f)(\f \g^{jk} \f). $$
No other term obtained by varying the structures in \C{fourfermi}\
contains a $5$ gamma piece. It is also easy to see that varying
the $g_2$ structure into $3$ fermions never gives a $5$ gamma
term. Lastly, the terms from varying $S_1$ contain either $a$ or
$\df$ and do not mix with this term. We can therefore conclude
that 
\be \label{vanishing} 
g_4^{(n,1)} = 0. \ee 
This obervation was already made for the case $n=2$ in~\cite{Hyun:1999hf}. Here
we see that it is true to all orders in the derivative expansion. 
The vanishing of this coupling implies
that the action is non-generic!

\subsection{Sources?}

While this coupling is absent for the homogeneous solution, it
might be generated by sources. To determine whether this is the
case, we need to compute the $N^i$ and $M$ terms inductively. 
 Fortunately, some of the required
computations have already been performed~\cite{Kazama:2002jm}. The generalization
of those results appears in Appendix~\ref{NM}\ for general
$n$. First note that in the absence of sources, the coefficient
functions for the homogeneous solution satisfy
the relations,  
\bea \label{coeffrel}
i g_2^{( n)} + 4 g_4^{(n,2)} - 2 \widetilde{g}_4^{(n,1)} &=& 0, \\
4  g_4^{(n,1)} + 4  g_4^{(n,3)} + {i \over r} \{ g_2^{( n)} \}' &=&
0, \\ \label{vanishingrel}
g_4^{(n,1)} &=& 0. 
\eea
Since the first few orders in the momentum expansion
are special, we will now proceed order by order.

\subsubsection{Terms of $O(v^4)$}

The first case is the $O(v^4)$ terms for which $n=2$. In this case,
$g_4^{(2,1)}$ must vanish since there are no sources,
\be
g_4^{(2,1)}(r) =0, 
\ee
 but we need to
determine $N^i_2$ and $M_2$. These terms are listed in
Appendix~\ref{NM}.  
What we want to know is whether any term in $N$ or $M$ can source
the coupling $g_4^{(3,1)}$ at $O(v^6)$. A quick perusal of the
expression~\C{NMfour}\ restricting to $n=2$ and setting
$g_4^{(2,1)}=0$ shows us that only one term in $M_2$ is relevant
\bea
M_2 \e & = & \ldots +  g_2^{( 2)} v^{2} ( x^i \f \g^{ij}\f)( \g^j
\e). 
\eea 
This term sources equation~\C{vanishingrel}\ via the variation of the
spin coupling,
$$ 2 g_2^{(2)} v^2 x^i v^j \f \g^{ij} M_2^{(2)}\e, $$
in $\dd^2 S_2$, which has a $5$-gamma piece that mixes with the
$g_4^{(3,1)}$ coupling at
$O(v^6)$. This is the only source with the right gamma matrix
structure in $\dd^2 S_2$. 
The homogeneous solution is therefore modified by this source,  
\be \label{vsix} 4 g_4^{(3,1)} - 2 \{g_2^{(2)} \}^2 = 0. \ee
The good news is that the previously vanishing spin-spin coupling is
induced at $O(v^6)$ by the $O(v^4)$
terms. Since the $O(v^4)$ terms are $1$-loop exact, this coupling is
$2$-loop exact, which is no great surprise since we know that all
couplings at $O(v^6)$ are $2$-loop exact. However, it is important to
note that this coupling is $2$-loop exact regardless of how the other
$O(v^6)$ couplings are renormalized. No independent argument is
required.  It should be possible to 
check the relation~\C{vsix}\ directly using the techniques
of~\cite{McArthur:1998gs}\ combined with the computations
of~\cite{Becker:1997wh, Becker:1997xw}.

\subsubsection{Terms of $O(v^6)$}

Let us now ask whether the spin-spin coupling
at $O(v^8)$ is similarly sourced. The potential sources are generated
in two ways: either from
contributions to the supersymmetry transformations generated at $O(v^6)$
which act on terms of $O(v^4)$, or from contributions generated at
$O(v^4)$ which act on terms of $O(v^6)$. Said differently, the sources
come from terms in the variations,
$$ \dd^2 S_3 + \dd^3 S_2. $$ 

To determine the contributions from $\dd^3 S_2$, we need to learn
about the terms in $N^{i(3)}$ and
$M^{(3)}$. which are generated at $O(v^6)$. First note that the homogeneous
solution for the coefficient
functions~\C{firsttwo}\ at $O(v^6)$ is now modified by the $O(v^4)$
sources, as we already saw in~\C{vsix}. Relation~\C{1ferm}\ becomes
\be \label{modcoeff}
g_2^{( 3)} = - {i \{ g_0^{( 3)} \}'\over 2 r} + {2i \over r} {d\over
  dr} \{
 g_0^{( 2)} \}^2.
\ee 
{}For the following argument, it turns out that we do not need the explicit form of
$N_0^{i(3)}$ and $M_0^{(3)}$ generated at $O(v^6)$. To see why we first need to realize
that these terms have
the same form as the homogeneous solution~\C{hom1} and~\C{hom2}, but
with different coefficient functions. This is not hard to show
explicitly, and is also true just on general grounds. So we need to
ask whether  an $(N_0^{i(3)}, M_0^{(3)})$ of this form can possibly mix with
the $g_4^{(4,1)}$ coupling by appearing in the variation $ \dd^3 S_2$.

This would have been the case had $g_4^{(2,1)}$ not vanished via
the $M_0^{(3)}$ to the supersymmetry transformations. 
Since this coupling does vanish, there is no mixing and we need
not worry about the explicit form of these terms. The same cannot be
said for $(N_2^{i(3)}, M_2^{(3)})$ which can, in principle, mix with
the  $g_4^{(4,1)}$
coupling.

Of the terms in the homogeneous solution appearing in~\C{NMfour}, only
two are relevant.
\bea \label{sour1}
M_2 \e & = & \ldots  - 2 i\widetilde{g}_4^{(3,1)} x^2 v^{4} x^i 
(\f \g^{ik}\f) (\g^k \e) 
+  g_2^{( 3)} v^{4} ( x^i \f \g^{ij}\f)( \g^j
\e). 
\eea 
Both these coefficient functions are modified by sources according
to~\C{vsix}\ and~\C{modcoeff}.
These terms acting on the $2$-fermion coupling at $O(v^4)$ can source
the spin-spin coupling at $O(v^8)$.  

Are there any relevant inhomogeneous pieces of $(N_2^{i(3)},
M_2^{(3)})$? These inhomogeneous terms are sourced from the
variation $\dd^2 S_2$. Let us examine these corrections term by
term. The first terms come, schematically, from the variation  
$$ \dd ( g_0^{(2)} v^4) $$
using $N_2^{i(2)}$. This variation has terms of the form $ x^i \e
N_2^{i(2)} \f$ or $ v^i \p_t (\e N_2^{i(2)} \f)$. A glance at the form
of $N_2^{i(2)}$~\C{NMfour} shows us that the new terms in  $(N_2^{i(3)},
M_2^{(3)})$ induced by this variation never have the right gamma
matrix structure
to source the spin-spin coupling. 

The same is true for the variation of schematic form 
$$ g_2^{(2)} \f^3
M_0^{(2)}\e $$ 
because $M_0^{(2)} \sim \g^k
v^k$, and $g_4^{(2,1)}=0$. It is less obvious, but also true, that the variation
$$ 2 g_2^{(2)} v^2 x^i v^j \f \g^{ij} M_2^{(2)} \e $$
gives rise to no relevant sources for $(N_2^{i(3)},
M_2^{(3)})$ although it does source the $g_4^{(3,1)}$ coupling. This
leaves the variation
$$ \dd ( g_2^{(2)} v^2 x^i v^j) \f \g^{ij} \f $$
using $N_0^{i(2)}$. Almost all the terms in this variation play no
role except for
$$ \ldots + g_2^{(2)}v^2 x^i \left\{ i g_0^{(2)} v^2 (\e \g^j \df)
\right\} (\f \g^{ij} \f). $$
This term sources $M^{(3)}$ giving a contribution,
\be \label{sour2}
M_2^{(3)} \e = \ldots - g_2^{(2)} g_0^{(2)} v^4 x^i (\g^j \e)(\f
\g^{ij}\f). 
\ee 
We now have all the ingredients needed to study the generation of the
3-loop spin-spin coupling at $O(v^8)$.

\subsection{A 3-loop prediction at $O(v^8)$}

Let us finally put together all the sources for $g_4^{(4,1)}$. The
$O(v^4)$ terms are generated at one-loop. By explicit
computation~\cite{Douglas:1997yp}, we know that
\be g_0^{(2)} = -{15\over 16} {1\over r^7}.
\ee
This generates $g_2^{(2)}$ via~\C{1ferm}. The other input we require
is the value of $g_3^{(0)}$. This involves knowledge of the
homogeneous solution at $O(v^6)$. In this particular case, we know
there is no independent homogeneous solution. All terms at $O(v^6)$
are determined by terms at $O(v^4)$. Again from an explicit
two-loop computation, we know that~\cite{Becker:1997wh}
\be
g_0^{(3)} = -{225\over 64} {1\over r^{14}}.
\ee
This again determines $g_2^{(3)}$ via~\C{modcoeff}. Lastly,
$g_4^{(3,1)}$ is determined by~\C{vsix}. All the unknown
functions are now fixed.

The 
first source contribution comes from $\dd^3 S_2$. The variation,
$$ 2 g_2^{(2)} v^2 x^i v^j \f \g^{ij} M_2^{(3)} \e, $$
acts as a source for $g_4^{(4,1)}$ when we consider the terms
$$  M_2^{(3)} = \ldots + \{ g_2^{(3)} - 2i x^2\widetilde{g}_4^{(3,1)} 
- g_2^{(2)} g_0^{(2)}
\} v^4 (x^i \f \g^{ij}\f)(\g^j \e). $$
There are no other sources from $S_2$. From $\dd^2 S_3$, we obtain a
similar contribution
$$ 2 g_2^{(3)} v^4 x^i v^j \f \g^{ij} M_2^{(2)} \e, $$
where the only relevant term involves
$$  M_2^{(2)} = \ldots + g_2^{(2)}v^2 (x^i \f \g^{ij}\f)(\g^j \e). $$
There is one other contribution from $\dd^2 S_3$ coming from 
$$ 4 g_4^{(3,1)} v^4 x_i x_j (\f \g^{ik} \f) (\f \g^{jk}
M_0^{(2)}\e), $$
where
$$ M_0^{(2)}\e = 3 g_0^{(2)} v^2 v^k \g^k \e. $$
Putting together all these contributions gives us the prediction, 
\be
4 g^{(4,1)} = 2 g_2^{(2)} \{g_2^{(3)} - 2i x^2\widetilde{g}_4^{(3,1)} 
- g_2^{(2)} g_0^{(2)} \} - 2 g_3^{(2)} g_2^{(2)}
- 12 g^{(3,1)} g_0^{(2)}.
\ee
More important than the specific numerical value is the claim that this
coupling is determined in terms of more relevant interactions, and that
the coupling is $3$-loop exact both perturbatively and non-perturbatively. 

\subsection{Generalizing the argument}

This argument generalizes in two ways. First, it should be clear
that by repeating the argument, one learns about the sources for the $(n-1)$-loop
the spin-spin coupling at $O(v^{2n})$. While we can predict that this coupling
is induced at this loop order by more relevant interactions, we need more
information to determine
the exact value of the coupling. At the moment, it is not sufficient
to know just the $O(v^4)$ terms. In particular, we need to know about
$g_2^{(m)}$ for all $m<n$. Or equivalently, we need to know about
$g_0^{(m)}$ for all $m<n$. The two couplings are related by~\C{1ferm}\
up to source terms. It might be the case that supersymmetry determines
all $(n-1)$-loop terms at $O(v^{2n})$ (in a way outlined in the
introduction), but that question is beyond the scope
of this analysis. What seems clear is that there will be more couplings
beyond the spin-spin coupling~\C{spincoupling}\ determined by more
relevant sources.

The second way the argument generalizes explains the existence of
rays in figure~\ref{fig}\ with slopes other than $1$. Most of our
previous discussion is not special to the $O(v^4)$ terms in any way. 
Suppose at $O(v^{2n})$, there is a new $m$-loop contribution to
$v^{2n}$. In terms of figure~\ref{fig}, this corresponds to an anchor
point at coordinates $(v^{2n}, m)$. By definition, nothing can source
these  terms so
we can conclude that
\be
g_4^{(n,1)} =0
\ee
at $m$-loops. However, a spin-spin coupling is generated at
$(m+1)$-loops at $O(v^{2n+2})$.  This comes about via the variation of the
spin coupling,
$$ 2 g_2^{(n)} v^{2n-2} x^i v^j \f \g^{ij} M_2^{(2)}\e, $$
using the $O(v^4)$ correction to the supersymmetry
transformations. So the coefficient of this induced spin-spin coupling
is completely determined. This is the reason a ray of slope $1$ emanates from
each anchor point. Indeed the same is true if we consider the
variation, 
$$ 2 g_2^{(n)} v^{2n-2} x^i v^j \f \g^{ij} M_2^{(k)}\e, $$
{}for any $k$. This is the reason that all possible rays (one for each
anchor point) emanate from
each anchor point, and the reason for the intricate web of induced
couplings depicted in figure~\ref{fig}.

\section*{Acknowledgements}

It is my pleasure to thank Eduard Antonyan, Sonia Paban and Mark Stern
for extensive discussions.
The work of S.~S. is supported in part by NSF CAREER Grant
No. PHY-0094328, and by the Alfred P. Sloan
Foundation.

\newpage
\appendix
\section{Fermion Structures}
\label{fermionstructures}
\subsection{Two fermion structures}

Using the
simplifications described in section~\ref{simplify}, we can determine
the form of the action~\C{notmom} completely. Each real fermion
transforms in the ${\bf 16}$ of
$Spin(9)$. We note that
$$ {\bf 16} \wedge {\bf 16} = [2] \oplus [3]$$
where $[n]$ refers to the antisymmetric $n$-form representation. The
basic fermion bilinears are therefore, 
$$ \f \g^{ij} \f, \quad \f \g^{ijk} \f. $$
We will call these $2$-gamma and $3$-gamma structures, respectively. 
All of the couplings in the Lagrangian are constructed from these
building blocks contracted with $x$ and $v$. For example, the only
possible $2$ fermion structure is 
\be \label{2fermion}
x^i v^j \f \g^{ij} \f.  \ee

\subsection{Four fermion structures}

To construct higher fermion structures, we will make use of some
simplifying identities. The basic Fierz identities found
in~\cite{Paban:1998ea, Hyun:1999hf}\
teach us that
\bea (\f \g^{ij} \f) (\f \g^{ij} \f) &=& 0, \cr
 (\f \g^{ijk} \f)( \f \g^{ijk} \f) &=& 0, \cr
(\f \g^{ij} \f)( \f \g^{ijk} \f) &=& 0, 
\eea 
and that $(\f \g^{ijk} \f)( \f \g^{imn} \f)$ can be expressed entirely
in terms of 2-gamma structures $ (\f \g^{pq} \f)$. When combined with
CPT invariance, which acts as complex conjugation while sending 
$$ x \rightarrow -x, \qquad t \rightarrow -t, $$
these constraints will allow us to express all $4$ fermion structures in
terms of 2-gamma structures alone. 

To see this, note that the $4$ fermion structure appears with a real
function of $(x,v)$ if we want a Hermitian coupling. It must therefore
be even in $x$, and by $Spin(9)$ invariance, even in $v$. The only
possible structure involving a $3$-gamma bilinear has the form, 
$$ (\f \g^{ijk} \f)(\f \g^{il} \f). $$
However, this will be odd in either $x$ or $v$. So we can restrict to
structures built from $2$-gamma bilinears. 

There are three possible terms
which all appear in the supersymmetric completion of $v^4$~\cite{}, 
\bea  
x_ix_j (\f \g^{ik} \f)(\f \g^{jk} \f), \cr 
v_i v_j (\f \g^{ik} \f)(\f \g^{jk} \f), \cr 
x_ix_j v_k v_l (\f \g^{ik} \f)(\f \g^{jl} \f). 
\eea
This is no surprise since the most general fermion structure
requires enough velocities so that we can, if we wish, attach a single velocity
to any fermion bilinear. With $2$ bilinears, this constraint means we
need 2 velocity factors which is precisely the number available at
order $n=2$.

\subsection{Beyond four fermions}

I cannot resist pushing this discussion a little further. How many $6$
fermion couplings exist? If we want a Hermitian term in the action
then each  $6$ fermion coupling appears with an imaginary
function of $x$ and $v$. 
Invariance under CPT then implies that the coupling is odd in
$x$. Lastly, these couplings appear with a $v^{2n-3}$ factor so the
coupling must also be odd in $v$. These are exactly the characteristics
enjoyed by the 2 fermion coupling~\C{2fermion}.  

It is not hard to check that any coupling constructed from a $3$-gamma
structure either fails to satisfy these constraints, or can be
rewritten in terms of $2$-gamma structures. The possible $2$-gamma
structures must be odd in both $x$ and $v$ so they can have $1$ or $3$
factors of $x$ or $v$. This gives the following $4$ possibilities (two
of the structures are simply $x\leftrightarrow v$ exchanges)
\bea
x_i v_j (\f \g^{ik} \f)(\f \g^{jl} \f)(\f \g^{kl} \f), \cr 
x_i x_j x_k v_p (\f \g^{ip} \f)(\f \g^{jl} \f)(\f \g^{kl} \f), \cr 
v_i v_j v_k x_p (\f \g^{ip} \f)(\f \g^{jl} \f)(\f \g^{kl} \f), \cr 
v_i v_j v_k x_p x_q x_r (\f \g^{ip} \f)(\f \g^{jq} \f)(\f \g^{kr} \f), 
\eea

How about $8$ fermion couplings? These couplings must be even in both
$x$ and $v$. Let us first consider the $3$-gamma structures. The only
way a  $3$-gamma structure can appear is in the combination 
$$  (\f \g^{ijk} \f)(\f \g^{il} \f). $$
Let us denote this structure by $(3+2)$. It has $3$ free indices. 
The first possible $3$-gamma structure is
$(3+2)+(2+2)$, but this combination has an odd number of free indices since
any contractions remove indices in pairs. It therefore cannot be even in
both $x$ and $v$, and is ruled out. The other possibility is
$(3+2)+(3+2)$. In this case, we are not allowed to contract the $3$-gamma
structures together since the result can be expressed in terms of
$2$-gamma structures, but there are other possible
contractions. It turns out, however, that all the other contractions
vanish. 

The remaining possibility is no contractions. In this case, we need to
consider the square of       
$$  x^j v^k  v^l (\f \g^{ijk} \f)(\f \g^{il} \f), $$
or the same structure with $x \leftrightarrow v$. This coupling does
not seem to vanish. It also does not seem reducible in an obvious way
to a product of $2$-gamma structures. 

The remaining possibilities are
built from $2$-gamma bilinears. A list of
structures contains
\bea \label{list4}
(\f \g^{ij} \f)(\f \g^{jk} \f)(\f \g^{kl} \f)(\f \g^{li} \f), \cr 
x^i x^m (\f \g^{ij} \f)(\f \g^{jk} \f)(\f \g^{kl} \f)(\f \g^{lm} \f), \cr
v^i v^m (\f \g^{ij} \f)(\f \g^{jk} \f)(\f \g^{kl} \f)(\f \g^{lm} \f),\cr
x^i x^k x^l x^m (\f \g^{ij} \f)(\f \g^{jk} \f)(\f \g^{lp} \f)(\f \g^{pm} \f), \cr
v^i v^k v^l v^m (\f \g^{ij} \f)(\f \g^{jk} \f)(\f \g^{lp} \f)(\f\g^{pm} \f), \cr
x^i x^k v^l v^m (\f \g^{ij} \f)(\f \g^{jk} \f)(\f \g^{lp} \f)(\f\g^{pm} \f), \cr
\{ x^j v^k  v^l (\f \g^{ijk} \f)(\f \g^{il} \f) \}^2 , \label{c1}\\
\{ v^j x^k  x^l (\f \g^{ijk} \f)(\f \g^{il} \f) \}^2. \label{can2}
\eea
Note that many of these structures are simply related by
$x\leftrightarrow v$. If the last $2$ structures of \C{list4}\ are
truly independent (as they appear to be) then they are likely to give rise to new
non-generic couplings in the action.  

It would be unpleasant if we now had to consider $10$ fermion
couplings and onwards. Fortunately, $m$ fermion couplings are related by Hodge
duality to $16-m$ fermion couplings, so we have already classified all
possible terms! To set conventions, let us define the Hodge dual  of
an $m$ fermion term, $T_{a_1\cdots a_m} \f_{a_1}
\ldots \f_{a_m}$ by
\be \ast T_{a_{m+1}\cdots a_{16}} = \epsilon_{a_1 \cdots a_m a_{m+1}
  a_{16}}  T^{a_1 \cdots  a_m}.
\ee

\newpage
\section{Corrections to the SUSY Transformations}
\label{NM}

In this Appendix, we will list the homogeneous form for $N_2^i$
and $M_2$ obtained by studying the
variation of the action \C{notmom}\ into $3$ fermions. This
generalizes the results appearing in~\cite{Kazama:2001sq}. By studying the terms
that involve no $a$ or $\df$, we see that
\bea 
i g_2^{( n)} + 4 g_4^{(n,2)} - 2 \widetilde{g}_4^{(n,1)} &=& 0, \\
4  g_4^{(n,1)} + 4  g_4^{(n,3)} + {i \over r} \{ g_2^{( n)} \}' &=&
0, \\
g_4^{(n,1)} &=& 0. 
\eea
By studying the $(a, \df)$ terms, we determine the homogeneous solution
for $N_2^i$ and $M_2$
\bea \label{NMfour}
\e N^i_2 \f &=& (4-4n) \widetilde{g}_4^{(n,1)} v^{2n-4} x^2 v^i
(x^l \f \g^{lk} \f)( \e \g^k \f) + \cr &&  (8-4n) \widetilde{g}_4^{(n,3)} v^{2n-6}
x^2 v^i  ( x^l v^k \f \g^{lk} \f)(v^p \e \g^p \f) - \cr && 
 2  \widetilde{g}_4^{(n,3)} x^2 v^{2n-4} (x^l \f \g^{li}\f)(v^p \e
\g^p \f)   - 2  \widetilde{g}_4^{(n,3)} x^2 v^{2n-4} (x^l v^k\f \g^{lk}\f)( \e
\g^i \f), 
\\
M_2 \e & =& - 2 i\widetilde{g}_4^{(n,1)} x^2 v^{2n-2} x^i \left[
  2(\g^{ik} \f)(\f \g^k \e) + (\f \g^{ik}\f) (\g^k \e) \right] - \cr &&
   2 i\widetilde{g}_4^{(n,3)} x^2 v^{2n-4} x^i v^k v^l \left[
  2(\g^{ik} \f)(\f \g^l \e) + (\f \g^{ik}\f) (\g^l \e) \right] + \cr &&
(2n-2) g_2^{( n)} v^{2n-4} ( x^i v^j \f \g^{ij}\f)( v^l \g^l \e) +
  g_2^{( n)} v^{2n-2} ( x^i \f \g^{ij}\f)( \g^j \e).
\eea 
This homogeneous solution will, in general, be modified by sources generated by more relevant
couplings in the effective action.

\bibliographystyle{amsunsrt-es}


\providecommand{\href}[2]{#2}\begingroup\raggedright\endgroup

\end{document}